# Soliton generation by counteracting gain-guiding and self-bending

Valery E. Lobanov,[1] Yaroslav V. Kartashov,[1,2] Victor A. Vysloukh,[1,3] and Lluis Torner[1]

[1]*ICFO-Institut de CienciesFotoniques, and UniversitatPolitecnica de Catalunya, 08860 Castelldefels (Barcelona), Spain*
[2]*Institute of Spectroscopy, Russian Academy of Sciences, Troitsk, Moscow Region, 142190, Russia*
[3]*Departamento de Fisica y Matematicas, Universidad de las Americas – Puebla, Santa Catarina Martir, 72820, Puebla, Mexico*
*Corresponding author: valery.lobanov@icfo.eu*

We introduce a new concept for stable spatial soliton formation, mediated by the competition between self-bending induced by a strongly asymmetric nonlocal nonlinearity and spatially localized gain superimposed on a wide pedestal with linear losses. When acting separately both effects seriously prevent stable localization of light, but under suitable conditions they counteract each other, forming robust soliton states that are attractors for a wide range of material and input light conditions.
*OCIS Codes: 190.4360, 190.6135*

Formation and stabilization of solitary waves is one of the central problems of nonlinear optics [1]. Solitary waves may form not only in conservative but also in dissipative settings where localization occurs by the exact balance between gain and losses, which may be spatially distributed [2,3]. Dissipative spatial solitons have been studied in different media, including lasers with saturable gain and absorption [2], systems where light evolution is governed by the cubic–quintic Ginzburg–Landau equation [3-9], semiconductor amplifiers [10], and settings with spatially localized gain and uniform nonlinear losses [11-17], to name just a few.

A common feature of the above systems is that solitons are supported by symmetric local nonlinearities and exist due to the competition between gain and losses arising from different physical origins. Thus, in systems described by the cubic-quintic Ginzburg-Landau equation solitons exist because of the exact balance between linear losses, cubic gain, and quintic losses [3-9]. In settings with spatially localized gain, stable states also form when local cubic losses compensate gain [11-17]. In contrast, localized gain acting in systems with focusing nonlinearity prevents formation of stable solitons in the absence of higher-order absorption. Analogously, if the nonlinearity of a uniform system is nonlocal and exhibits an asymmetric component, a self-induced transverse drift takes place and non-accelerating stationary states cannot form.

In this Letter we show that the above two effects, that seriously prevent stable light localization when acting separately, may also counteract each other and thus generate stable attractors. The setting analyzed here is simple, as it includes spatially shaped *linear* gain and losses combined with only a *conservative focusing* nonlinearity that includes an asymmetric nonlocal component, such as the diffusive nonlinearity exhibited by photorefractive crystals [18-21]. In contrast to conventional schemes for dissipative soliton formation (where linear or nonlinear gain is compensated by higher-order absorption [2-17]) in our case stable solitons with asymmetric shapes may form only when self-bending, caused by the asymmetric nonlocality, drives the beam out of the gain-domain into the region with linear losses, thereby affording a stable balance between gain, attenuation, nonlinearity and diffraction. To the best of our knowledge, this is first ever example where asymmetry of a nonlocal nonlinear response is a key ingredient for the existence of stable soliton states.

We address the propagation of a laser beam along the $\xi$-axis of a medium with localized gain, homogeneous linear losses, symmetric local and asymmetric nonlocal components of the focusing nonlinear response, that can be described by the nonlinear Schrödinger equation for the dimensionless light field amplitude $q$:

$$i\frac{\partial q}{\partial \xi} = -\frac{1}{2}\frac{\partial^2 q}{\partial \eta^2} - q|q|^2 + \mu q \frac{\partial}{\partial \eta}|q|^2 + i\gamma(\eta)q. \qquad (1)$$

Here $\eta$ and $\xi$ are normalized transverse and longitudinal coordinates, respectively; the function $\gamma(\eta)$ describes the gain profile, and $\mu$ is the self-bending parameter. For $\mu \neq 0$ the nonlinear contribution to the refractive index becomes an asymmetric function of the transverse coordinate even for symmetric inputs. We consider a Gaussian gain profile $\gamma(\eta) = a\exp(-\eta^2/d^2) - \gamma_0$, superimposed on a background of linear losses $\gamma_0 > 0$. Such gain landscapes may be realized, for example, by using transverse optical pumping of doped planar photorefractive waveguides. We use values for the self-bending parameter $\mu \sim 0.2$, which correspond to light beams with width of the order of $3\ \mu\text{m}$ at the wavelength $\lambda = 0.5\ \mu\text{m}$, propagating in a photorefractive material biased with a static electric field $\sim 400\ \text{V/cm}$ and maintained at room temperature. A value $a = 1$ corresponds to the gain coefficient $\sim 75\ \text{cm}^{-1}$. In our calculations we set $\gamma_0 = 1$ and $d = 1$.

First, we considered the case of a *symmetric* focusing nonlinearity, i.e., $\mu = 0$. In such a case, soliton solutions can be found for gain values below a threshold value $a = a_0$ (for $d = 1$ one has $a_0 \approx 1.639$) for which a localized mode exists even in the linear limit supported by the gain-guiding effect [22]. However, in media with focusing local nonlinearity such solutions are completely unstable in their entire existence domain $(a < a_0)$, because increasing the energy flow results in a higher concentration of light in amplifying domains and thus in an exponential growth of the peak amplitude. Thus, elucidation of physical mechanisms that support stable states in *focusing* media with spatially shaped *linear* gain and losses remains a challenge.

We thus consider media with an *asymmetric* component of the nonlocal nonlinear response described by Eq. (1) with $\mu \neq 0$. We use Gaussian beams as input light conditions and we model their propagation for large distances by integrating numerically Eq. (1). Figure 1 shows the typical propagation dynamics we observed for different values of the self-bending parameter $\mu$. The plot illustrates the central result of this Letter: stable solitons can form within a range of values of the amplification $a$ and self-bending $\mu$ parameters, despite the focusing character of the nonlinearity. Figure 1(b) depicts an illustrative example. Within their stability domain, the generated solitons are attractors. They have asymmetric profiles, shifted from the axes of the amplifying domain. Outside the stability domain, beams undergo decay or formation of pulsating structures, which emit radiation along the direction dictated by the sign of the parameter $\mu$ [Fig. 1(a)]. Such pulsations occur because the almost exponential growth of the intensity within the amplifying domain results in a rapid increase of the transverse intensity gradients and, consequently, in a remarkable enhancement of the self-bending effect. Such enhancement causes emission of radiation from the amplifying domain into the absorbing one, leading to a quasi-periodic decrease of the energy flow of the beam and thus reduction of the self-bending effect. In a certain parameter range, such emission becomes continuous instead of pulsed and hence a steady state is achieved.

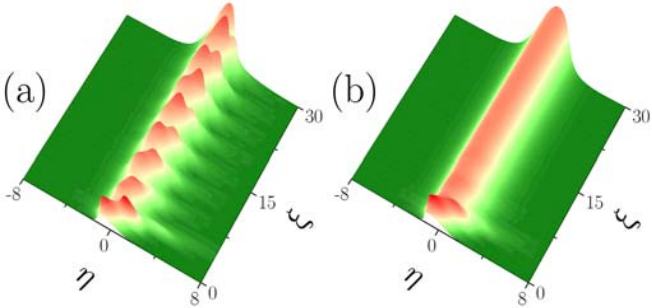

Fig. 1. Propagation of Gaussian beam in a medium with localized gain for $a=2$ at $\mu=0.2$ (a) and $\mu=0.3$ (b).

Therefore, guided by Fig. 1, we searched for solutions in the form $q(\eta,\xi)=w(\eta)\exp(ib\xi)$, where $b$ is the propagation constant, the function $w=w_r+iw_i$ describes the shape of the beam, and $w_r, w_i$ are the real and imaginary parts of the complex amplitude $w$. We searched for solutions by using a relaxation method, using the gain-loss balance condition $\int_{-\infty}^{\infty}\gamma(\eta)|w(\eta)|^2 d\eta=0$ and the fact that in dissipative media all parameters describing a stationary state are set by the gain landscape. Stability of the solutions was analyzed upon substitution of the perturbed field $q=[w_r+iw_i+(u+iv)\exp(\delta\xi)]\exp(ib\xi)$, where $u,v$ are small perturbations with the complex growth rate $\delta=\delta_r+i\delta_i$, into Eq. (1), linearization of the resulting expressions and solving the corresponding eigenvalue problem. Solutions are linearly stable as long as $\delta_r \leq 0$.

In the presence of self-bending ($\mu \neq 0$) and for gain values $a<a_0$ two completely new soliton branches appear, which exist for $\mu<\mu_{\text{upp}}$, i.e. if self-bending is not too strong [see Fig. 2(a) that shows the dependence of soliton energy flow $U=\int_{-\infty}^{\infty}|w(\eta)|^2 d\eta$ on $\mu$ for different gain parameters $a$]. The two branches merge at the point $\mu=\mu_{\text{upp}}$.

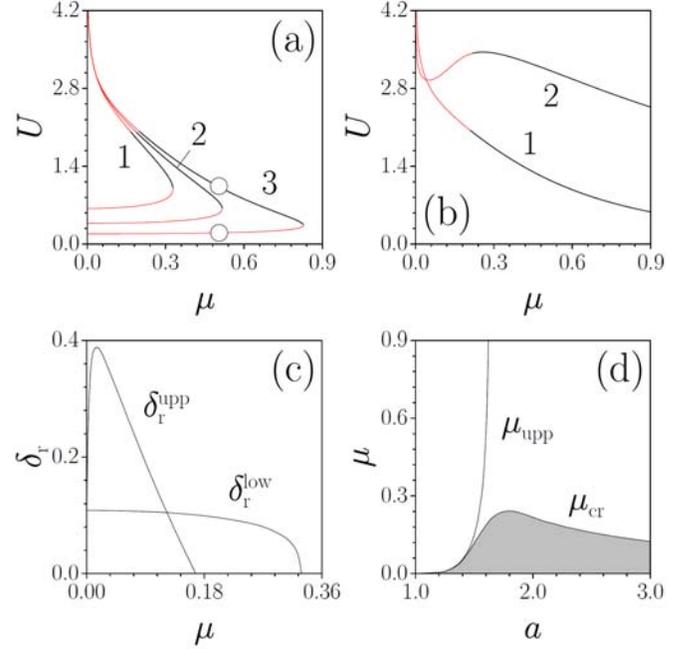

Fig. 2. (a) $U$ versus $\mu$ at $a=1.57$ (curve 1), $a=1.6$ (curve 2), and $a=1.62$ (curve 3). Circles correspond to the fields shown in Figs. 3(a) and 3(b). (b) $U$ versus $\mu$ at $a=1.64$ (curve 1) and $a=2$ (curve 2). In (a) and (b) stable branches are shown black. (c) $\delta_r$ versus $\mu$ for upper $\delta_r^{\text{upp}}$ and lower $\delta_r^{\text{low}}$ soliton branches at $a=1.57$. (d) Domains of soliton existence and soliton stability in the plane . Solitons are unstable in the shaded area for $\mu<\mu_{\text{cr}}$ and stable in the white area for $\mu_{\text{cr}}<\mu<\mu_{\text{upp}}$.

Spatially localized solutions belonging to the lower branch are unstable and rapidly decay or transform into the solution from the upper branch in the course of propagation. In contrast, a considerable part of the upper branch at $\mu>\mu_{\text{cr}}$ ($\delta_r \to 0$ as $\mu \to \mu_{\text{cr}}$) corresponds to stable solitons, as predicted by the linear stability analysis and verified by direct propagation of the perturbed solutions. Fig. 2(c) depicts the real part of the corresponding perturbation growth rate versus $\mu$ for the lower $\delta_r^{\text{low}}$ and upper $\delta_r^{\text{upp}}$ branches. It is inside such stability domain, namely $\mu_{\text{cr}}<\mu<\mu_{\text{upp}}$, that excitation of stable solitons with asymmetric shapes is possible. The upper border of the existence domain in $\mu$ rapidly increases and diverges as $a \to a_0$ [Fig. 2(d)]. For gain coefficients exceeding $a_0$, the energy flow becomes a single-valued function of the self-bending parameter $\mu$, i.e. only one branch survives, and the existence domain becomes infinitely wide in terms of $\mu$ [see Fig. 2(b) for representative $U(\mu)$ dependencies]. Solutions from such single branch were found to be stable for $\mu>\mu_{\text{cr}}$.

The domains of existence and stability of the solutions obtained in media with asymmetric nonlocality and localized gain are shown in Fig. 2(d). Note the rapid growth of the width of the stability domain with increasing gain. Importantly, stabilization may be achieved even for small values of the self-bending parameter, i.e., for $\mu \sim 0.2$.

The stable soliton solutions exhibit a nontrivial phase distribution. Figure 3 shows typical soliton profiles. While solutions from the lower branch at $a < a_0$ are nearly symmetric and centered around $\eta = 0$ [Fig. 3(a)], those from the upper branch feature notably asymmetric profiles [Fig. 3(b)] and their peak amplitudes are shifted from the center of the gain landscape. Both, the asymmetry of the soliton profile and the spatial shift increase when the gain grows or when the parameter $\mu$ decreases [Fig. 3(c)].

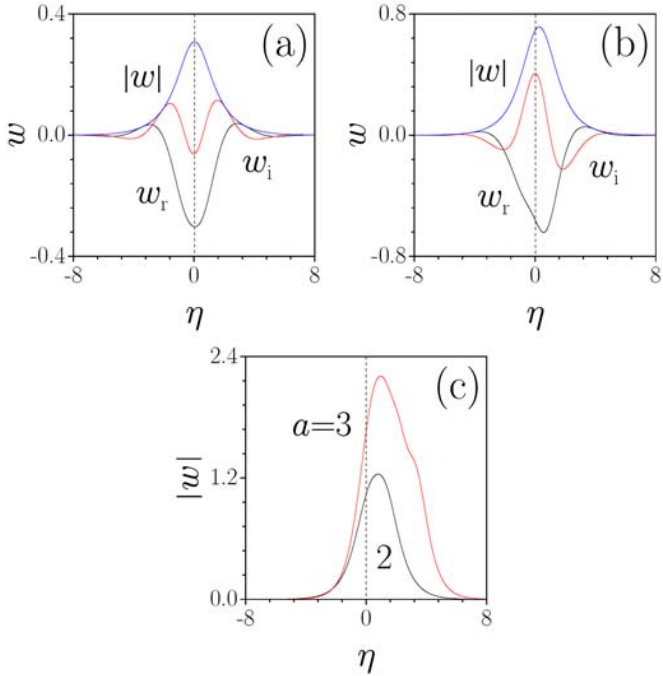

Fig. 3. Profiles of solutions at $\mu = 0.5$, $a = 1.62$ from the lower (a) and upper (b) branches. (c) Field modulus distributions at $\mu = 0.2$. Dashed lines indicate the center of the amplifying channel.

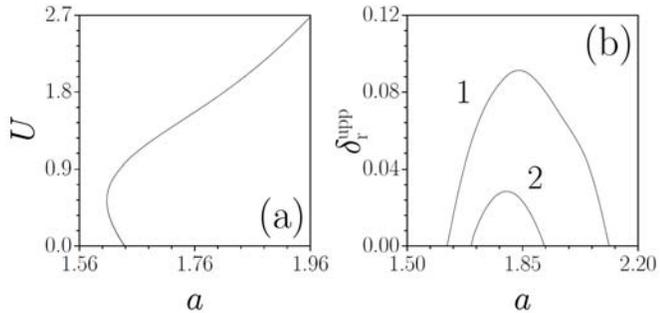

Fig. 4. (a) $U$ versus $a$ at $\mu = 0.6$. (b) $\delta_r$ for the upper soliton branch versus $a$ at $\mu = 0.2$ (curve 1) and $\mu = 0.23$ (curve 2).

For a fixed value of the $\mu$ parameter, we found that solutions exist above a threshold gain value [Fig. 4(a)]. Solutions from the lower branch are unstable, while those from the upper one are stable in a suitable domain [the dependence $\delta_r^{\text{upp}}(a)$ is shown in Fig. 4(b)]. At the upper branch the energy flow is a monotonically growing function of $a$, while at the lower branch $U$ decreases with increasing gain and vanishes when $a \to a_0$. For small values of the self-bending coefficient solutions were found to be unstable. However, the instability domain shrinks with growing $\mu$ [Fig. 4(b)], to an extent that at $\mu > 0.25$ all solutions from the upper branch are stable.

To elucidate the robustness of the trapping and stabilization mechanism afforded by the counterbalance between self-bending and gain-guiding, we verified that the results reported here remain qualitatively similar for different widths of the amplifying domain $d$. We found that the critical gain parameter $a_0$ at which the upper boundary of the existence domain tends to infinity, decreases with increasing the width of the amplifying domain (for example, at $d = 2$, $a_0 \approx 1.284$, at $d = 1$, $a_0 \approx 1.639$, and at $d = 0.5$, $a_0 \approx 2.539$). In addition, we verified that the shapes of the stability and existence domains are not sensitive to the exact value of $d$. Even the critical value of the self-bending coefficient, at which stabilization is achieved at large gain values (i.e., $\mu_{\text{cr}}$), does not change greatly when $d$ changes.

In summary, we introduced a new mechanism for spatial soliton formation and stabilization, based on the competition between self-bending effects in media with a strongly asymmetric nonlocal nonlinearity and spatially localized gain, superimposed on a pedestal of linear losses. The stability domain for such solitons, in terms of gain and self-bending parameters, is remarkably wide.